\documentclass[%
 aip,
amsmath,amssymb,
 reprint,%
]{revtex4-1}

\usepackage{graphicx}
\usepackage{dcolumn}
\usepackage{bm}
\renewcommand{\selectlanguage}[1]{}
\usepackage{tabularx}
\usepackage[utf8]{inputenc}
\usepackage[T1]{fontenc}
\usepackage{mathptmx}
\usepackage{etoolbox}
\usepackage[caption=false,font=normalsize,labelfont=sf,textfont=sf]{subfig}
\usepackage{floatrow}
\usepackage{amsmath,amsfonts}
\makeatletter
\def\@email#1#2{%
 \endgroup
 \patchcmd{\titleblock@produce}
  {\frontmatter@RRAPformat}
  {\frontmatter@RRAPformat{\produce@RRAP{*#1\href{mailto:#2}{#2}}}\frontmatter@RRAPformat}
  {}{}
}%
\makeatother

\begin{document}

\preprint{AIP/123-QED}

\title{Modelling Realistic Multi-layer devices for superconducting quantum electronic circuits}

\author{Giuseppe Colletta}
\affiliation{ 
James Watt School of Engineering, University of Glasgow, Glasgow G12 8QQ, UK 
}
\email{g.colletta.1@research.gla.ac.uk}

\author{Susan Johny}%
\affiliation{ 
James Watt School of Engineering, University of Glasgow, Glasgow G12 8QQ, UK 
}
\affiliation{National Physical Laboratory, Hampton Road, Middlesex, Teddington TW11 0LW, UK}
\author{Jonathan A. Collins}
\affiliation{ 
James Watt School of Engineering, University of Glasgow, Glasgow G12 8QQ, UK 
}
\author{Alessandro Casaburi}
\affiliation{ 
James Watt School of Engineering, University of Glasgow, Glasgow G12 8QQ, UK 
}
\author{Martin Weides}
\affiliation{ 
James Watt School of Engineering, University of Glasgow, Glasgow G12 8QQ, UK 
}


\date{\today}

\begin{abstract}
In this work, we present a numerical model specifically designed for 3D multilayer devices, with a focus on nanobridge junctions and coplanar waveguides. Unlike existing numerical models, ours does not approximate the physical layout or limit the number of constituent materials, providing a more accurate and flexible design tool. We calculate critical currents, current phase relationships, and the energy gap where relevant. We validate our model by comparing it with published data. Through our analysis, we found that using multilayer films significantly enhances control over these quantities. For nanobridge junctions in particular, multilayer structures improve qubit anharmonicity compared to monolayer junctions, offering a substantial advantage for qubit performance. For coated multilayer microwave circuits it allows for better studies of the proximity effect, including their effective kinetic inductance.

\end{abstract}
\maketitle

Most superconducting quantum technologies rely on Josephson Junctions \cite{RevLuk79, RevGol2004,intrtrasmon} 
and coplanar waveguides (CPW) \cite{CPW1,wallraffcpw} for their operation. To meet the rigorous standards required for such applications, the simulation of these components becomes an essential part of the fabrication process. 
This work develops a numerical model tailored for coplanar waveguides (CPW) \cite{CPW1,wallraffcpw} and nanobridge junctions, which offer a promising alternative to traditional tunnel junctions\cite{devoret2004superconducting,hertzberg2021laser} in superconducting quantum devices, such as qubits \cite{Arute2019,Miyajima:18} and Single Flux Quantum (SFQ) \cite{tolpygo2018superconductor} due to several key advantages. Unlike tunnel junctions, which rely on an oxide layer that can degrade performance by creating localized states, energy loss, and reduced coherence \cite{Burnett2019, PhysRevLettMartin}, nanobridges use a small metallic bridge to connect electrodes directly. This structure removes the oxide layer entirely, minimizing unwanted capacitance that interferes with precise qubit engineering and eliminating fabrication complexities that often hinder integration with other on-chip components \cite{collins}. These advantages make nanobridges particularly suitable for superconducting qubits. Previous theoretical works were limited to 3-D simulations\cite{3dFEMSNS} that did not include applications in quantum technologies or 2-D nanobridge designs \cite{Vijayjunction, Ragazzidikupr_2021, MiniaturizationofJosephsonJunctions, 2023_Kup_Gol} which did not consider more elaborate shapes such as rounded edges, which have an effect on the junction properties as it will be shown later in the paper. Therefore, we aim to overcome these limitations and interface our model with the standard design process of superconducting quantum devices. We also explore additional scenarios where our model can be applied, such as calculating the effect of the proximity effect in interfaces between superconductors and other metals, to establish the reduction of the energy gap in the structure which can have a detrimental effect on qubits' coherence time. This area has attracted enormous interest, especially since recent studies have shown that surface encapsulation techniques, involving passivating the surface of superconducting materials like niobium to prevent the formation of lossy oxides, can significantly enhance qubits' performance \cite{encapsulation}. This broader applicability underscores the versatility of our model in enhancing the design and performance of sophisticated superconducting quantum devices, paving the way for more reliable and scalable quantum technologies.

\begin{figure*}[t]
\centering
\subfloat[]
{\includegraphics[width=1.5in]{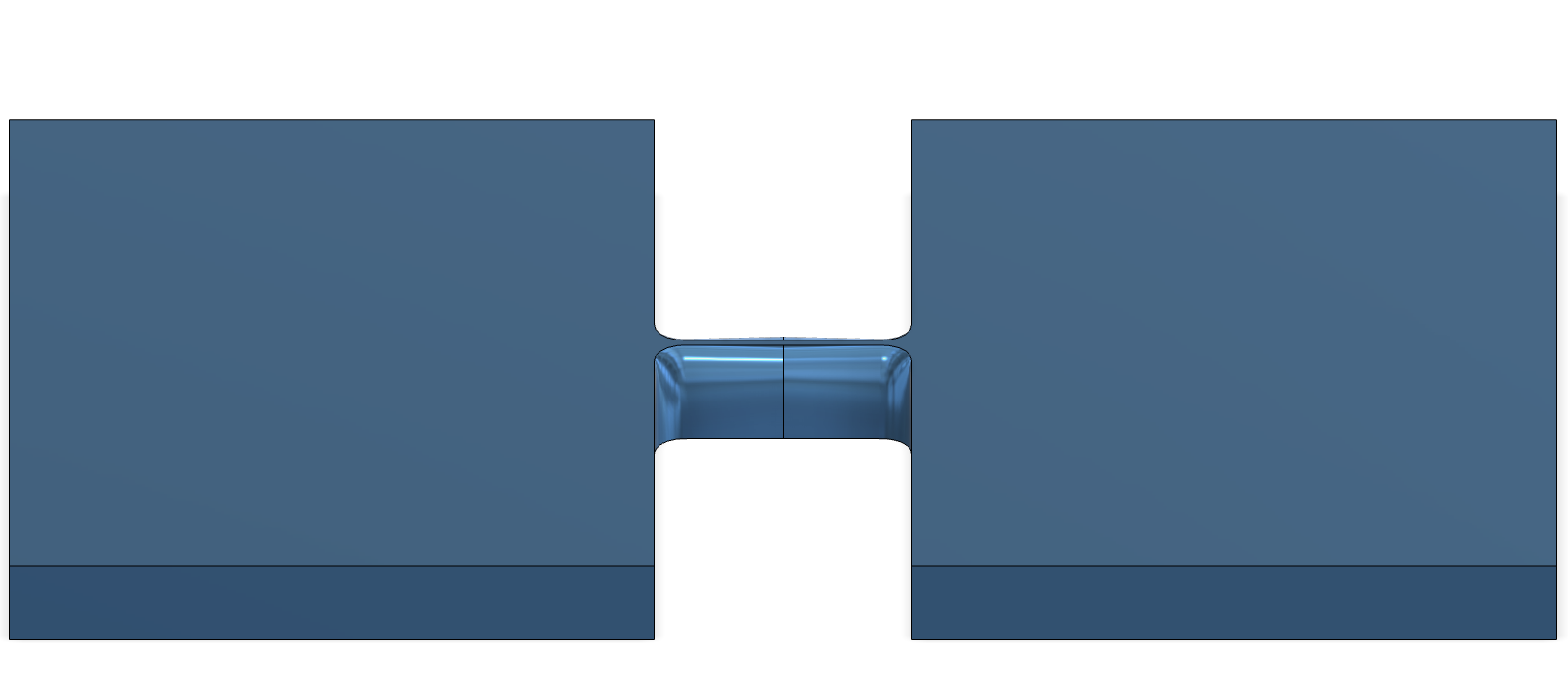}} \quad 
\subfloat[]
{\includegraphics[width=1.5in]{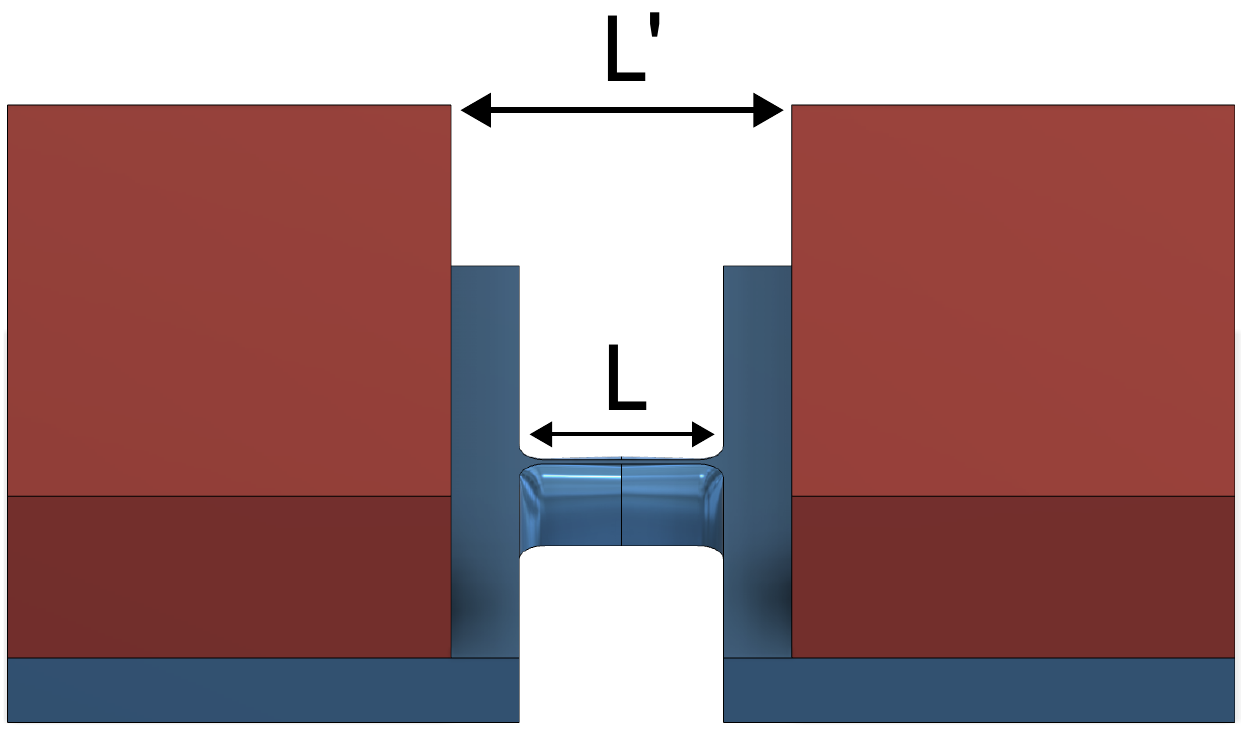}} \quad
\subfloat[]
{\includegraphics[width=1.5in]{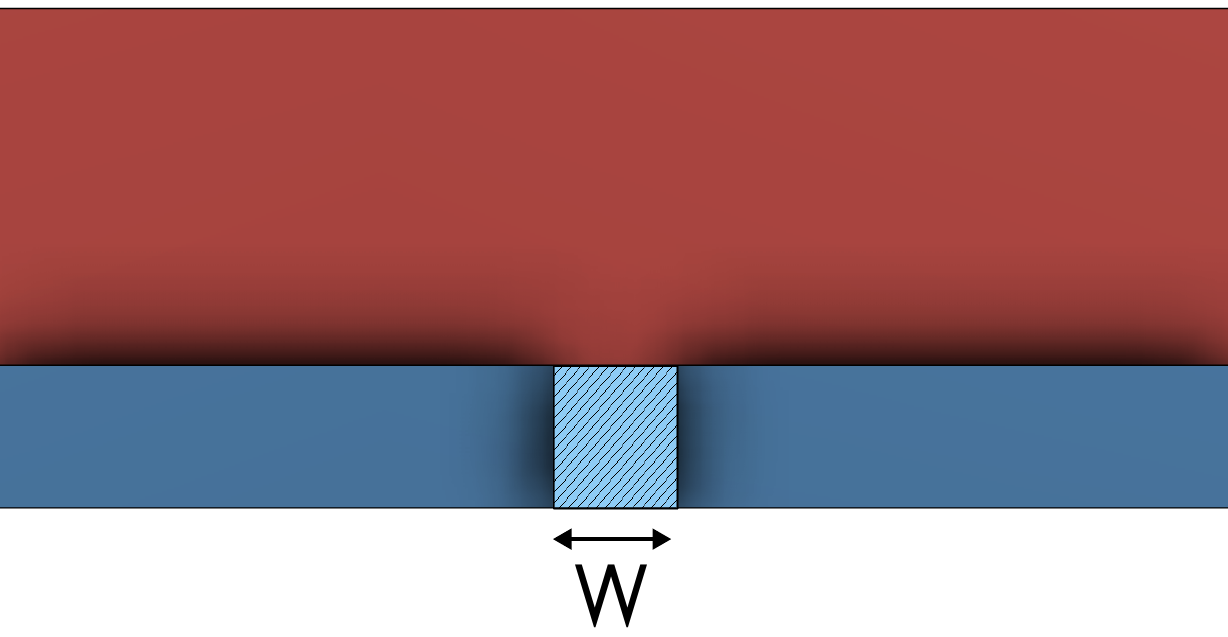}} \quad 
\subfloat[]
{\includegraphics[width=1.56in]{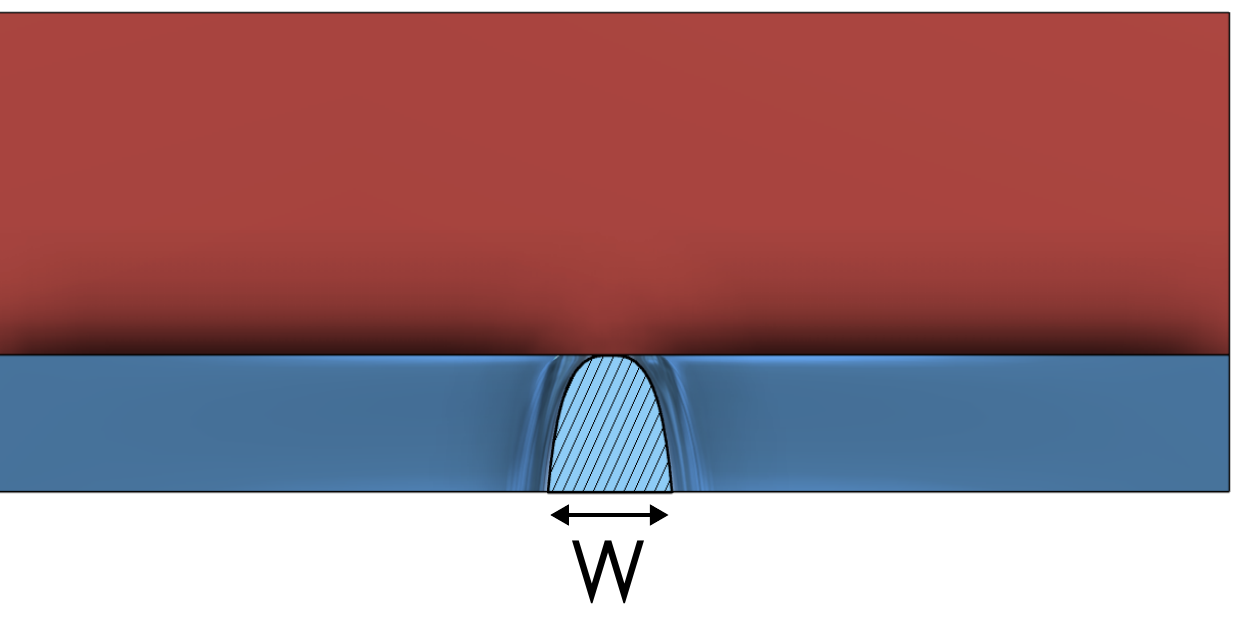}} \quad
\caption{3D design of the nanobridge junctions that will be analyzed later in the manuscript. (a) Planar Nanobridge junction  (b) Variable-thickness Nanobridge (VTB) junction, (c) cross-section of simplified VTB junction with rectangular edges and (d) cross-section of VTB junction with realistically rounded edges. Colours are to indicate the presence of different materials}
\label{fig:designnano1}
\end{figure*}

A nanobridge Josephson junction consists of two superconducting electrodes connected by a small constriction, usually of order of a few tens of nanometers. These constrictions may consist of the same material or a different one, with the latter referred to as SS'S or SNS junctions, where S represents a superconductor and S' or N represents the filling material that connects the two electrodes, which may be another superconductor (S') or a normal metal (N).


Under the assumption that the magnetic effects are negligible and all materials satisfy the dirty limit condition ($ l \ll \xi $), where $l$ is the mean free path, and $ \xi$ is the superconducting coherence length, we can describe the current transport properties of this type of junction with the Usadel equations \cite{RevGol2004, Usadel70, Kurpianov1988, Ragazzidikupr_2021}.   

\begin{equation}
-i D_I \nabla \cdot \hat{G}_I \nabla \hat{G}_I + [\tau_3 E + i \hat{\Delta}, \hat{G}_I] = 0.
\end{equation}

Here, \( \hat{G}_I \) is the \( 2 \times 2 \) matrix Usadel Green’s function in layer \( I \). The term \( E \) represents the energy, and \( D_I = v_F l_I / 3 \) is the diffusion coefficient in layer \( I \), where \( v_F \) is the Fermi velocity. The term \( \tau_3 \) is the third Pauli matrix, and \( \hat{\Delta} \) is the pair potential matrix, defined as  

\begin{equation}
\tau_3 =  
\begin{pmatrix} 
1 & 0 \\ 
0 & -1 
\end{pmatrix}, 
\quad 
\hat{\Delta} =  
\begin{pmatrix} 
0 & \Delta_I \\ 
\Delta_I^* & 0 
\end{pmatrix}.
\end{equation}
The coherence length, the diffusion coefficients and the critical temperature ($T_C$) of a superconductor are related by
\begin{equation}
    \xi_I=\sqrt{\hbar D_I/2 k_B \pi T_{C}}
\end{equation}
To determine stationary properties of the junction, such as current densities, the Usadel equations are solved for imaginary energies \( E_n = -i \omega_n \), where the Matsubara frequencies are given by \( \omega_n = \pi T (2n+1) \). In this regime, it is convenient to parametrize the Green’s function \( \hat{G} \) using the \(\Phi\)-parametrization:
\begin{equation}
\label{eq:Usadel}
\Phi_I(\omega_n, \vec r)=\Delta_I (\vec r) + \xi_i^2 \frac{\pi T_{CS}}{\omega_n G_I}\vec \nabla \cdot (G_I^2\vec\nabla \Phi_I(\omega_n, \vec r))
\end{equation}
which must be solved for $\Phi_I$ and $\Delta_I$ using the self-consistent equation:
\begin{equation}
\label{eq:selfcons}
\Delta_I(\vec r) \ln{\frac{T}{T_{CI}}}+2\pi T \sum^N_{n=0}{\bigg(\frac{\Delta_I(\vec r)}{\omega_n}-\frac{G_I\Phi_I(\omega_n, \vec r)}{\omega_n}\bigg)}=0,
\end{equation}
and the boundary conditions between different materials:
\begin{equation}
\label{eq:boundaryN}
\gamma_B \xi_{S'/N} G_{S'/N}\nabla\Phi_{S'/N}\cdot\vec{n} = G_S(\Phi_S-\Phi_{S'/N})
\end{equation}
\begin{equation}
\label{eq:boundaryN2}
\gamma_B \xi_S G_S\nabla\Phi_S\cdot\vec{n} = \gamma G_{S'/N}(\Phi_{S'/N}-\Phi_S)
\end{equation}
where $\vec{n}$ is the normal at the boundary,$T_{CS}$ represents the critical temperature of the superconducting electrodes, $k_B$ is the Boltzmann constant and $G_I=\omega_n/\sqrt{\omega_n^2+\Phi^*\Phi}$.
Defining $\rho_I$ as the resistivity of material $I$, $\gamma=\rho_S\xi_S/\rho_{S',N}\xi_{S',N}$ and $\gamma_B=R_B/\rho_{S',N}\xi_{S',N}$ represent the suppression parameter and the boundary resistance parameter, respectively. The suppression parameter describes how much the superconductivity is reduced going from S to the weak link (no suppression means $\gamma=0$), while the boundary resistance parameter describes the effect of the surface resistance ($R_B$) at the boundary between $S$ and $S'/N$. 
Far from the bridge on the $S$ surfaces not connected to the $S'/N$ layer, this boundary condition is applied:
\begin{equation}
\label{eq:boundaryext}
    \Phi_{S} = \Delta_0 e^{\pm i\delta/2}
\end{equation}
where $\Delta_0=1.76k_B \cdot T_{CS}$ is the Bardeen–Cooper–Schrieffer (BCS) energy gap and $\delta$ is the phase difference between the bulk electrodes \cite{Ragazzidikupr_2021}.
The superconducting current is calculated as follows:
\begin{equation*}
I_S \rho_N = \frac{\pi T A}{e} \sum^N_{n=0} Im\{G_I^2\Phi_I^*\partial_x\Phi_I\}
\end{equation*}
where $\rho_N$ is the resistivity of the whole junction in the normal state and $A$ is the cross-sectional area.

We will, now, discuss how we reformulated and solved the set of PDE in Eqs.(\ref{eq:Usadel}, \ref{eq:selfcons}).
The set of PDE is not only non-linearly dependent on the parameter $\Phi$ but also non-differentiable in $\mathbb{C}$, because of the term $\|\Phi\|^2$.
This added difficulty forced us to separate the real and imaginary parts of the equations to unsure convergence with the Newton method, actually creating a set of coupled equations that in the weak form become:
\begin{multline*}
	\int_\Omega G_S\operatorname{Re}\{\Phi_S(\omega_n)\}\cdot s_r\, d\Omega - \int _\Omega G_S \Delta_S \cdot s_r\, d\Omega + \\
	\xi_S^2 \frac{\pi T_{CS}}{\omega_n}\int_\Omega G_S^2\vec\nabla \operatorname{Re}\{\Phi_S(\omega_n)\}\vec \nabla \cdot s_r \, d\Omega +\\ \xi_S\frac{\pi T_{CS}}{\omega_n}\int_{\Gamma_{S(S'/N)}} \frac{\gamma G_S G_{S'/N}}{\gamma_B} \big(\operatorname{Re}\{\Phi_S(\omega_n)\}-\\ \operatorname{Re}\{\Phi_{S'/N}(\omega_n)\}\big)\cdot s_r\, d\Gamma_{S(S'/N)}= 0
\end{multline*}
and for the imaginary part:

\begin{multline*}
\int_\Omega G_S\operatorname{Im}\{\Phi_S(\omega_n)\}\cdot s_i\, d\Omega - \int _\Omega G_S \Delta_S \cdot s_i\, d\Omega + \\
\xi_S^2 \frac{\pi T_{CS}}{\omega_n}\int_\Omega G_S^2\vec\nabla \operatorname{Im}\{\Phi_S(\omega_n)\}\vec \nabla \cdot s_i \, d\Omega +\\ \xi_S\frac{\pi T_{CS}}{\omega_n}\int_{\Gamma_{S(S'/N)}} \frac{\gamma G_S G_{S'/N}}{\gamma_B} \big(\operatorname{Im}\{\Phi_S(\omega_n)\}-\\ \operatorname{Im}\{\Phi_{S'/N}(\omega_n)\}\big)\cdot s_i\, d\Gamma_{S(S'/N)}= 0
\end{multline*}
Here, $s_{r}$ and  $s_{i}$ are the test function sassociated respectively to $\operatorname{Re}\{\Phi_S(\omega_n)\}$  and $\operatorname{Im}\{\Phi_S(\omega_n)\} $ \cite{FEM}, $\Gamma_{S(S'/N)}$ represents the surface separating the materials $S$ and $(S'/N)$. The equations for the second material, denoted as $(S'/N)$, are identical except for differences in the integrals over $\Gamma_{S(S'/N)}$ due to distinct boundary conditions as outlined in Eq.(\ref{eq:boundaryN}).
The equations were solved in sfepy \cite{sfepy}, by creating new customs non-linear functions.

\floatsetup[figure]{style=plain,subcapbesideposition=top}
\begin{figure}[h]
\centering

\sidesubfloat[]
{\includegraphics[width=3.0 in]{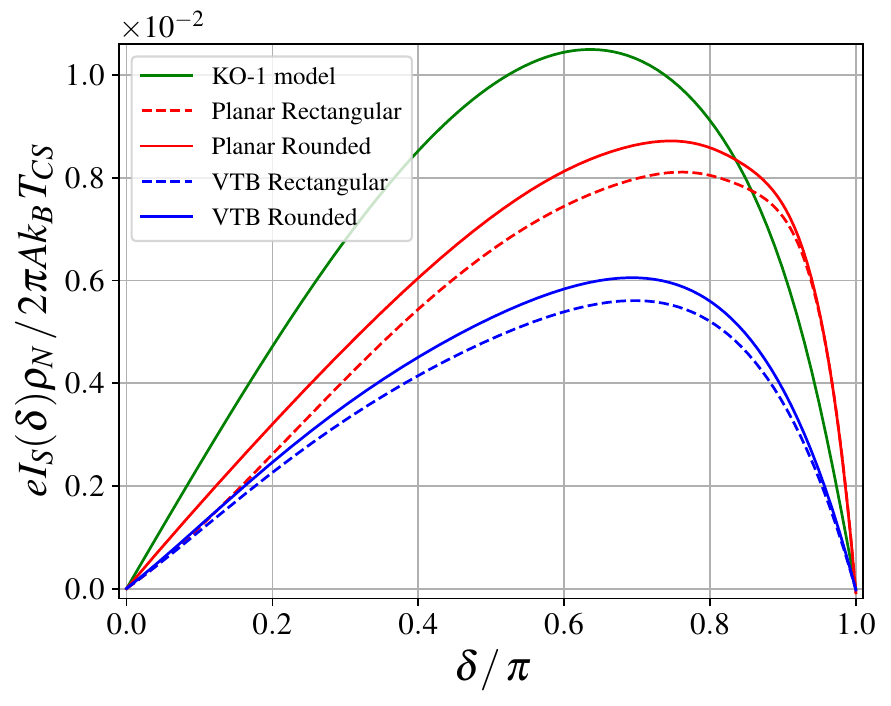}
\begin{picture}(0,0)
\put(-120,30){\includegraphics[height=1.8cm]{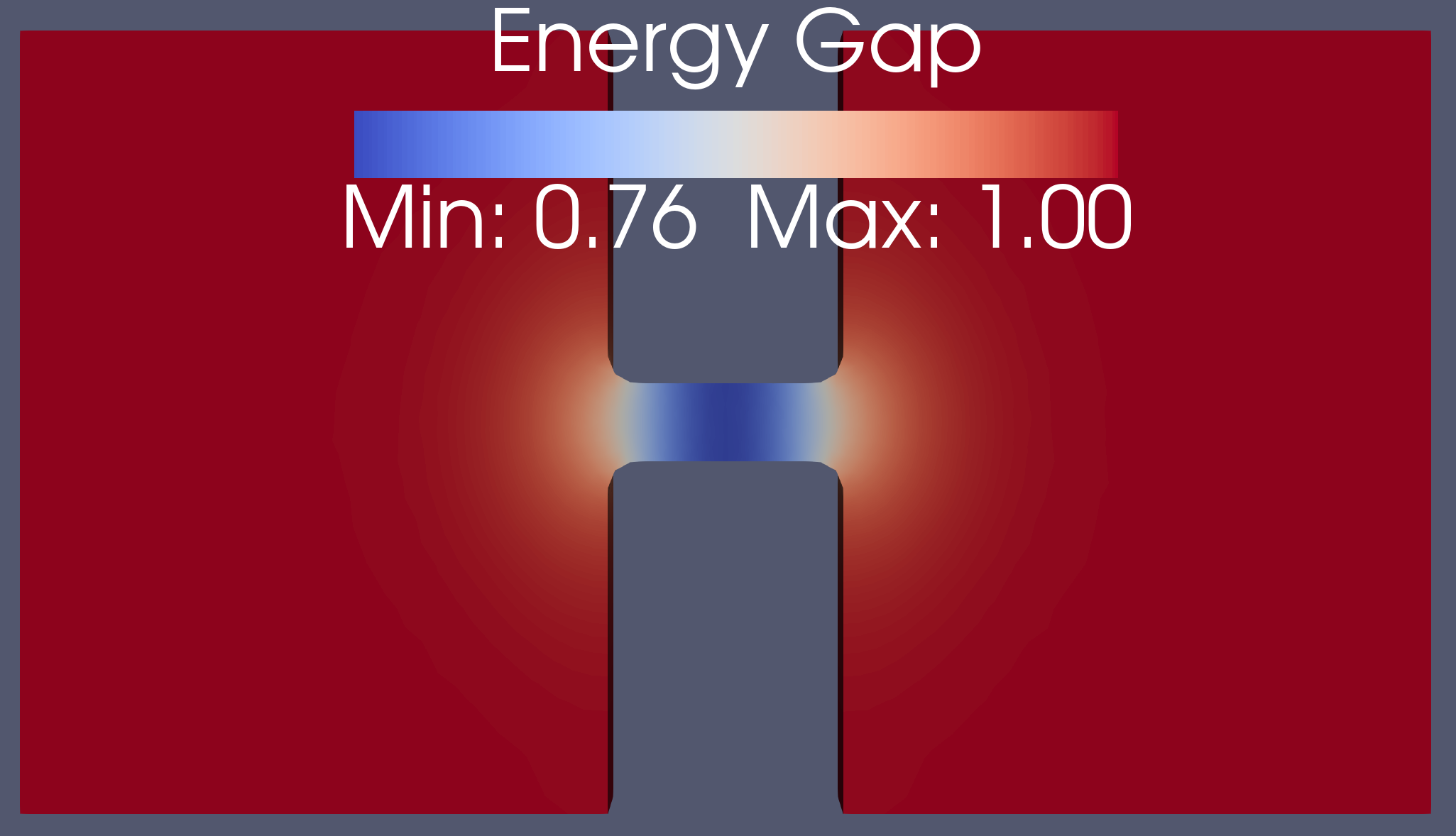}}
\end{picture}}\\
\sidesubfloat[]
{\includegraphics[width=3. in]{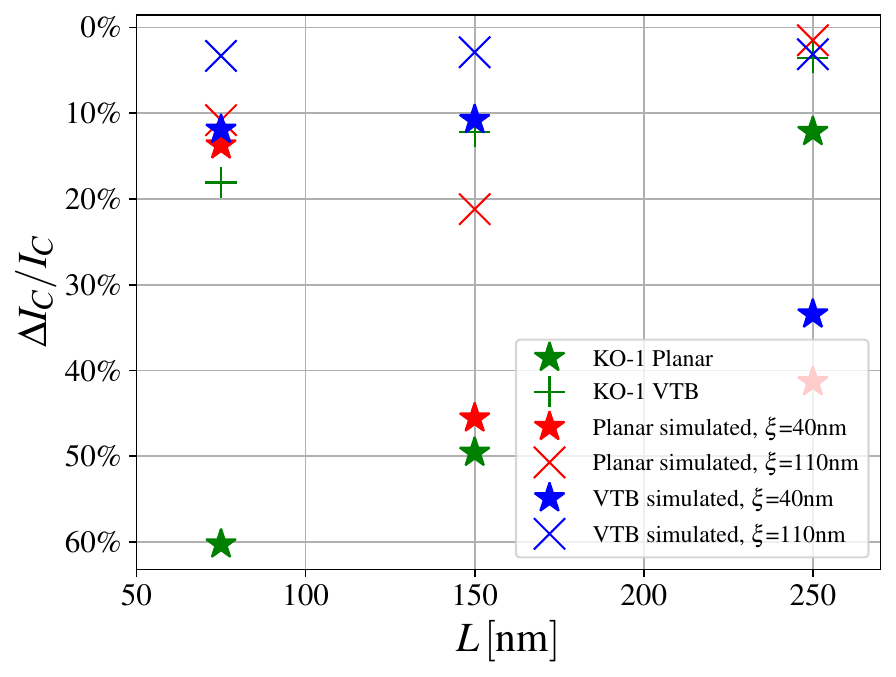}} \\
\caption{(a) Simulated CPR of different Nanobridge Junctions for $L/\xi_{S'}=2$, 
 $L'=2L$, $T/T_{CS'}=0.1$, $T_{CS}/T_{CS'}=5$, $\xi_S=5\xi_{S'}$, $\gamma_b=0.01$, $\gamma = 0.01$, $T_{CS}=5 K$ for the case of VTB. For the planar junction, there is only one material so the parameters are  $L/\xi_{S}=2$, $T/T_{CS}=0.1$, $T_{CS}=5 K$. $A$ is the cross-sectional area of the nanobridge. In the inset, we show the normalized pair potential of the "planar rounded" Junction (simulated with the parameters aforementioned and $\delta=0.3\pi$), as expected it reduces in the weak link. (b) Comparison between simulated and experimental\cite{VijaySquid} critical currents at different nanobridges' lengths, specifically we show the difference in percentage between the values reported in the paper and the one obtained either with the simulations or the KO-1 model. The simulated data agree to a good extent with the experimental ones, the alignment is further improved if a higher coherence length than the one reported in the paper is used \cite{propertiesAl}. }
\label{fig:simulationCPR1}
\end{figure}

\begin{figure} [b!]
\centering
\sidesubfloat[]
{\includegraphics[width=3.05 in]{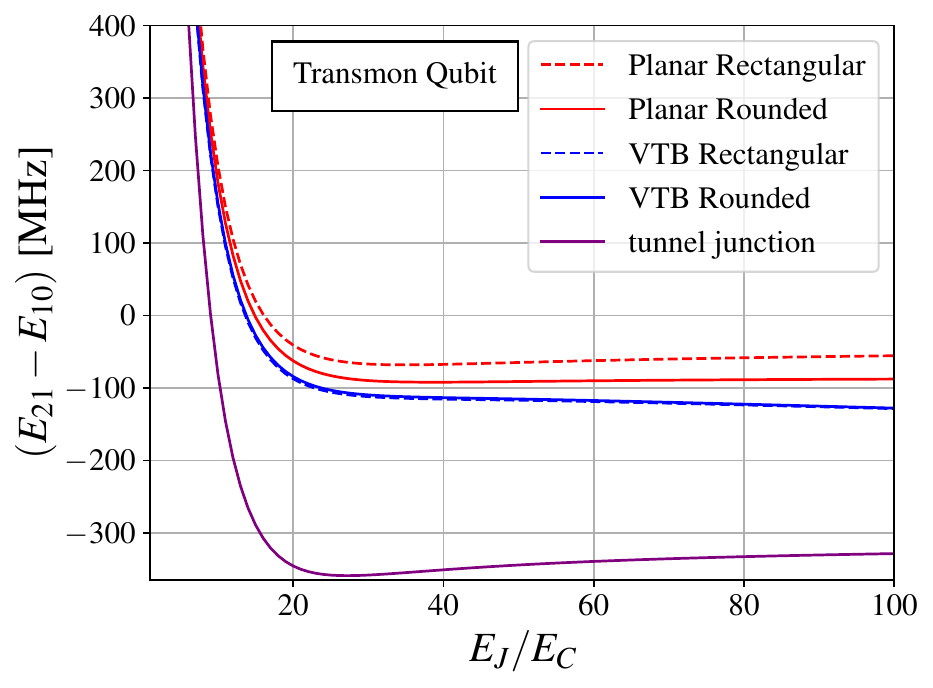}} \quad
\sidesubfloat[]
{\includegraphics[width=3.05in]{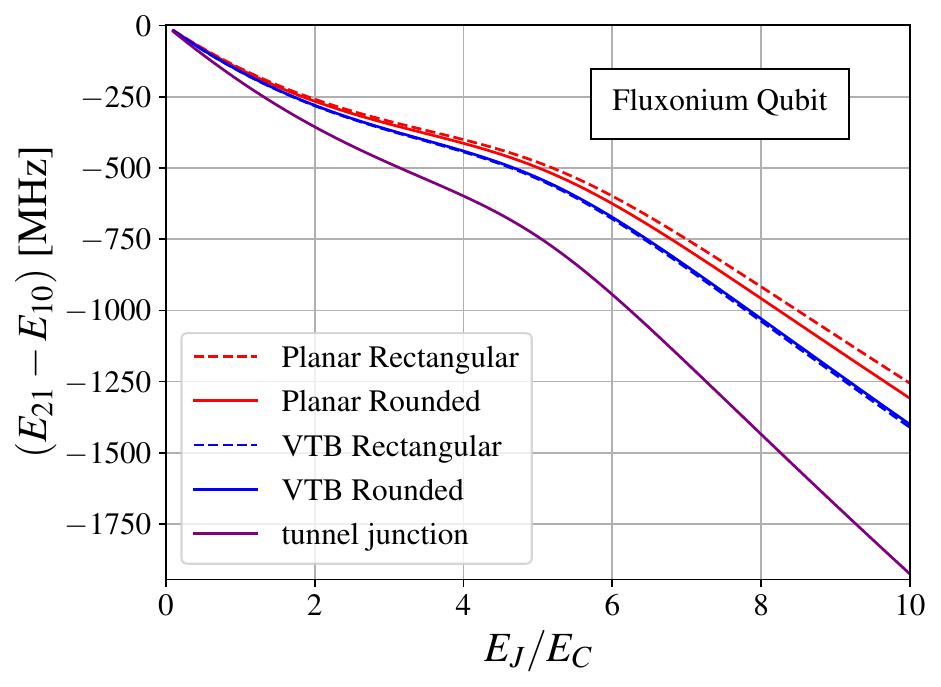}} \\
\caption{Simulated Transmon (a) and Fluxonium (b) qubit anharmonicities with the CPR in Fig. \ref{fig:simulationCPR1} supposing the same Josephson energy for all of them. $E_C= 300$ MHz was used for both qubits and $E_L=0.53 E_C$\cite{Fluxonium} was used for the Fluxonium. In the latter case smaller values of $E_J$ were explored due to the fact that Fluxonium qubit works in the regime $E_J<10 E_C$\cite{FluxoniumwithEjvalue}.}
\label{fig:simulationqubitene}
\end{figure}

\begin{figure*}[!ht]
\centering
\subfloat[]
{\includegraphics[width=4.72in]{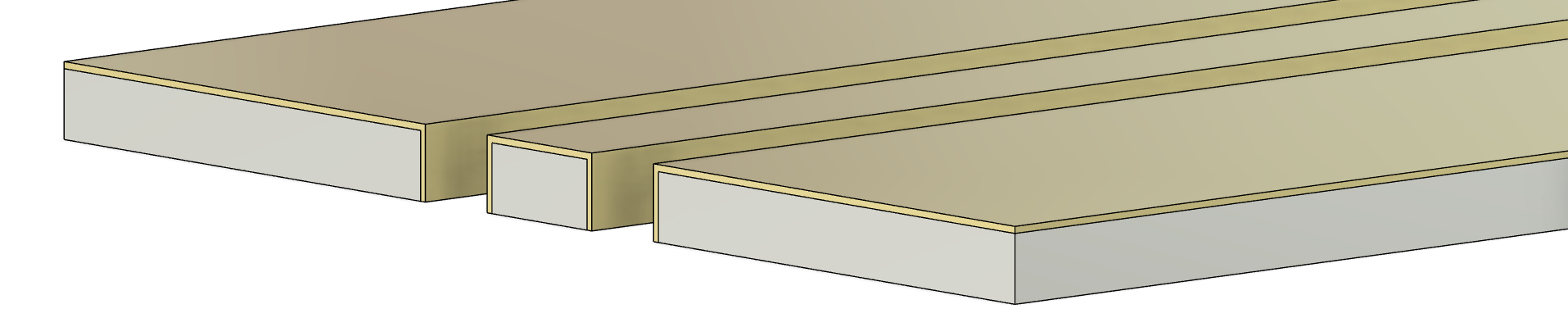}} \\
\subfloat[]
{\includegraphics[width=2.2in]{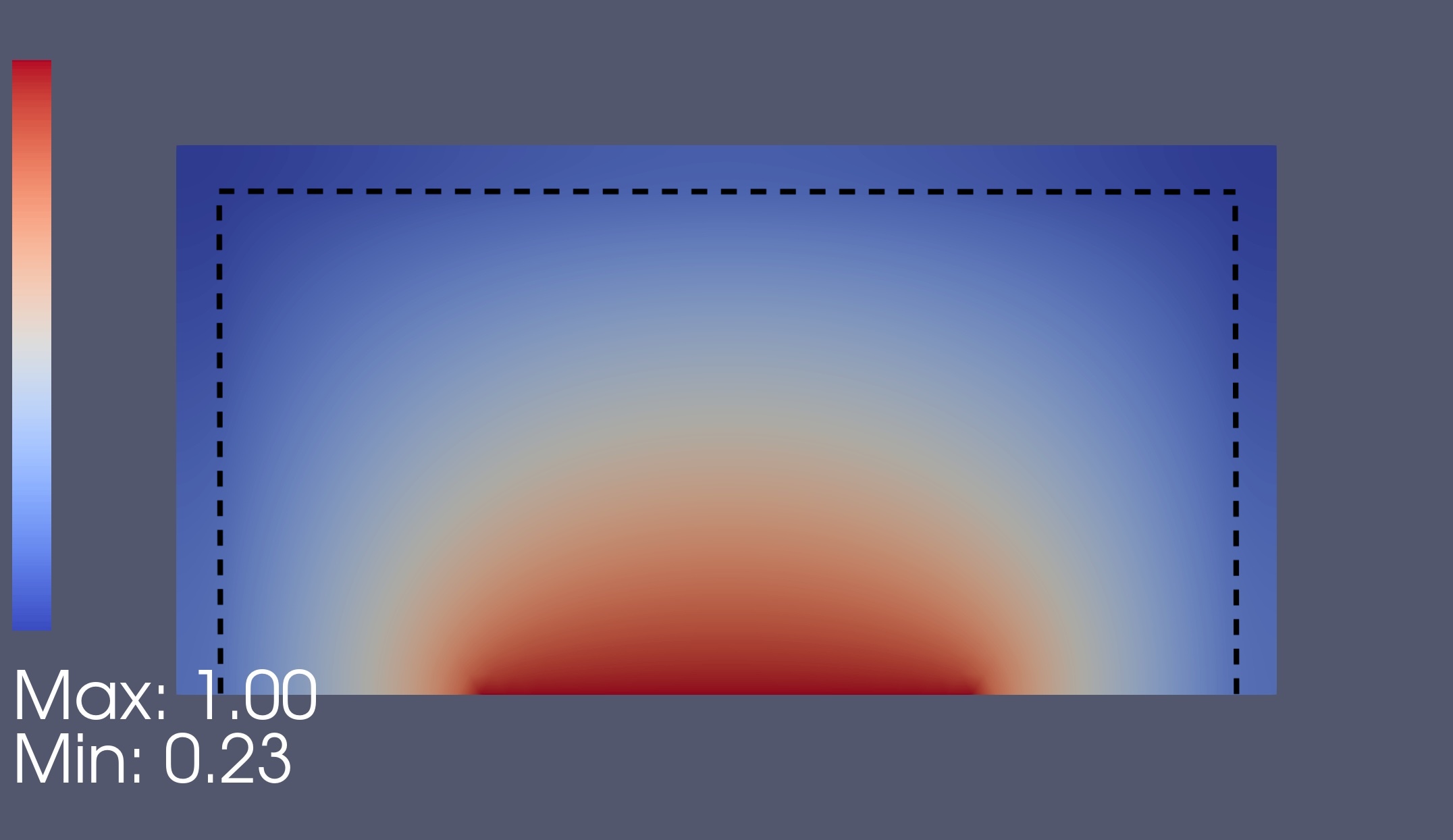}} \quad 
\subfloat[]
{\includegraphics[width=2.2in]{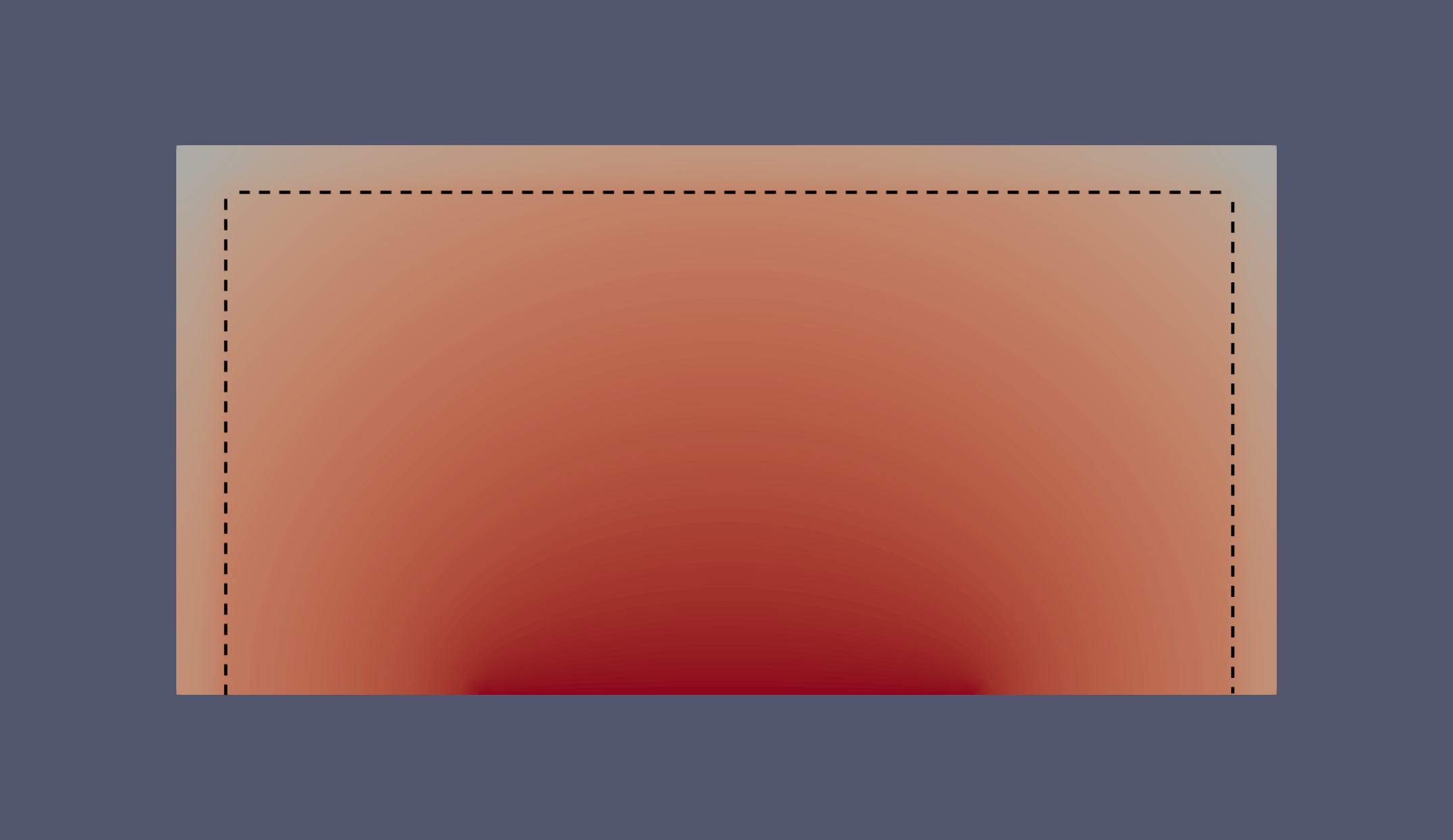}} \quad
\subfloat[]
{\includegraphics[width=2.2in]{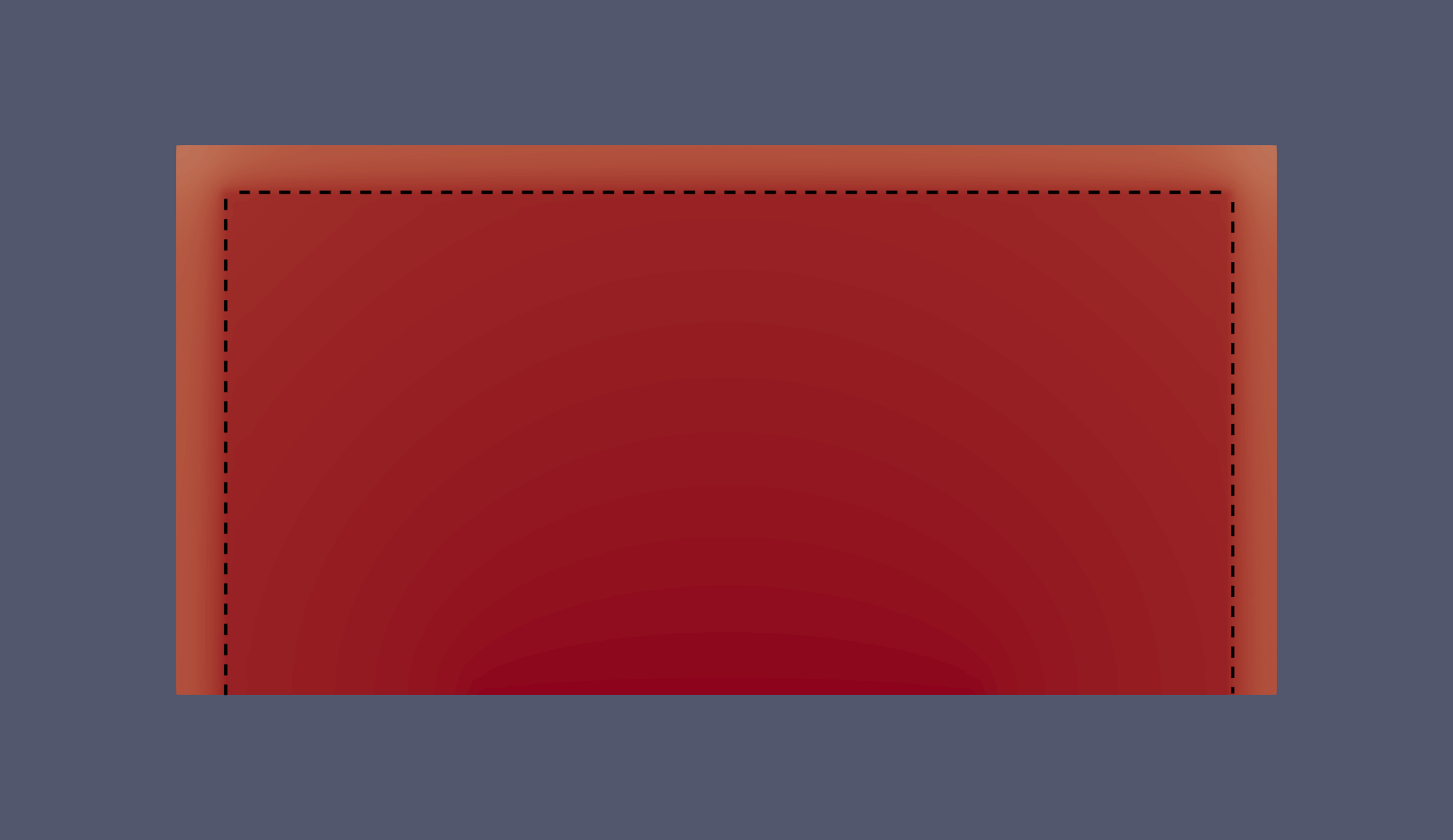}} \quad 
\caption{ (a) 3D design (not to scale) of a multilayer CPW. Magnitude of function $\Phi$ in the CPW for (b) $\gamma=10$ (c) $\gamma=1$ (d) $\gamma=0.1$. The dotted line was added to the images to indicate the separation between the niobium and encapsulating layer. Other parameters used were $\xi_S=5\xi_{N}$, $\xi_{N}=30 nm$, $\gamma_b=0.01$, $T/T_{CS}=0.1$, and the thickness of the superconducting and normal metal is 200 and 20 nm respectively. It is evident that the proximity effect is highly sensitive to the value of $\gamma$. As this parameter increases, the magnitude of function $\Phi$ reduces even when the thickness of the encapsulating material is smaller than the coherence length.}
\label{fig:res}
\end{figure*}

Superconducting quantum computing circuits require Josephson junctions with small critical currents, of the order of 10 to 100 nA, and CPRs as close as possible to a sinusoid in order to obtain non-linear energy levels with high anharmonicity\cite{nanobridgequbits}. For SFQs the constrain of low critical currents is lifted but the shape of the CPR still plays a crucial role in the generation of the pulses \cite{RSFQ}. Moreover, to stay within the Josephson effect, the dimensions of the bridge should be less than 3 times the coherence length \cite{RevGol2004}. For these reasons, different layouts have been numerically investigated as shown in Fig. \ref{fig:designnano1}, also taking into account realistically rounded edges that can come for an eventual etching process \cite{collins}, the lateral dimension W of the nano-bridge was chosen to be $W=0.5\xi$ and $L=2\xi$ for all the simulated devices.  
We calculated the current-phase relationship (CPR) for the following use-case scenarios: \begin{itemize} \item Planar junctions with rectangular and rounded cross-sections \item Variable Thickness Nanobridge SS'S Junctions (VTB) with rectangular and rounded cross-sections \end{itemize} These devices are illustrated in Fig.\ref{fig:designnano1}. The simulation results are presented in Fig.\ref{fig:simulationCPR1}, where we demonstrate how the rounded edges reduce the critical current in the junction, even after normalizing for the reduced cross-section. Additionally, in the case of the planar junction, the rounded edges slightly shift the maximum of the CPR to the left. This occurs because a reduced cross-section weakens the connection between the electrodes while maintaining the $L/\xi$ ratio, bringing the junction closer to the ideal case described by the one-dimensional Usadel equations, known as the KO-1 model \cite{KUPRIYANOV1981}, which is also shown in the figure.
Compared to the KO-1 model, the CPR exhibits lower values and increased skewness, which can be attributed to the reduction of the energy gap in the bridge area relative to the bulk electrodes, as illustrated in the inset. This reduction does not occur in the KO-1 model, where only the bridge is simulated with the assumption that the energy gap at its boundaries matches the bulk value. However, VTB junctions, for the same set of parameters, produce less skewed CPRs, which can improve qubit anharmonicity as discussed previously. This improvement is primarily due to the variable thickness design, where the electrodes act as a robust phase reservoir, leading to a more significant phase change across the bridge rather than within the electrodes, compared to planar junctions. This result aligns with the findings in Ref. \onlinecite{Vijayjunction}. The use of different materials enables better control over the device's critical current $I_C$ by selecting a bottom material with the appropriate resistivity.
We further validated our model by comparing it with the experimental work in Ref. \onlinecite{VijaySquid}, where aluminium VTB and planar nanobridges of different lengths were measured, as shown in Fig. \ref{fig:simulationCPR1}
(b). To do this, we first estimated the resistivity from the IV curve of a 75 nm long VTB nanobridge, and then plotted the simulated data obtained by using both the reported coherence length of 40 nm and an adjusted value of 110 nm, found in Ref. \onlinecite{propertiesAl}, which aligned more closely to the experimental data. This adjustment is based on the assumption that the coherence length might be larger than the reported one, calculated by using the film's resistivity that could have been made at room temperature, thus lowering the value of $\xi$. This comparison underscores the accuracy of our model and validates our design choices. 

\noindent
We are now using our numerical model to optimize the calculation of energy levels in superconducting qubits. The Hamiltonian for the transmon qubit with a junction featuring a generic current-phase relation (CPR) is given by:
\begin{equation}
\label{eq
}
\hat H = 4E_C(\hat n - n_g)^2 - E_J f(\hat\varphi),
\end{equation}
while the Hamiltonian for the fluxonium qubit is \cite{fluxoniumtrue, Fluxonium, FluxoniumwithEjvalue}:
\begin{equation}
\label{eq
}
\hat H = 4E_C(\hat n - n_g)^2 - E_J f(\hat\varphi) + \frac{E_L}{2}\varphi^2.
\end{equation}
In the case of the fluxonium, the following constraints need to be imposed to differentiate it from other types of inductively shunted devices \cite{FluxoniumwithEjvalue}, $E_L \ll E_J$ and $E_J < 10 E_C$, which means that the critical current of the nanobridge junction in these devices should be limited to less than 50 nA. Although this is a small value for a nanobridge junction, it can be achieved by utilizing a highly resistive layer.
From the CPR in Fig.\ref{fig:simulationCPR1} we calculated the associated Josephson energy as \cite{nanobridgequbits}:
\begin{multline*}
E_{J,1} f(\hat\varphi_1)= \int I_J(\varphi) V dt=\\ \int I_J(\varphi) \frac{\Phi_0}{2\pi}\frac{d\varphi}{dt}dt= 
\int I_J(\varphi) \frac{\Phi_0}{2\pi}d\varphi
\end{multline*}
To solve the equation we incorporated in scqubits \cite{Scqubits} and qiskit metal \cite{qiskit} a custom Hamiltonian in the phase basis and solved the eigenvalue problem with scipy.linalg.eigh \cite{SciPy} following the approach in Ref. \onlinecite{nanobridgequbits}.
The eigenenergies of the Hamiltonian are solved for different realistic ratios of $E_j/E_C$ and then used to calculate the anharmonicities plotted in Fig.\ref{fig:simulationqubitene}. As expected the anharmonicity reduces going from a tunnel junction (sinusoidal CPR) to the CPR we obtained with our model; this aligns with previous results \cite{nanobridgequbits, zlatkoCPRqubit}. However, the low levels of planar Junctions' anharmonicity are recovered using a multilayer design thanks to its less skewed CPR.

\setlength{\tabcolsep}{0.5em}
\begin{table*}[t]
    \centering
    {\renewcommand{\arraystretch}{1.5}
    \begin{tabularx}{0.9\textwidth}{c c c c c c c c c}
    \hline
    \hline
    Material &thickness [nm]& $ \rho [\mu\Omega \cdot$cm] & $\xi $ [nm] &$\gamma$ &$T_C$ [K]& $\text{min}(\Delta_{cap})/\Delta_0$ [$ \% $]& $    \big(L_{k.cap}/L_{k,0}\big)$ &References\\
    \hline
      Au& 10   & 20 & 88 & 0.16 & 0 & 95 & 0.97&  Refs. \onlinecite{meanfreepathGold, fewmetalselectronpath}\\
    TiN &15   & 300 & 53.6 &0.02 & 4.5 & 99 &0.99& Refs. \onlinecite{TiNnanobridge,TiNproperties,ALDTiN,propertiesTiN} \\

      Al &10   & 1.5 & 110 &1.69& 1.2 & 97 &0.69& Ref. \onlinecite{propertiesAl} \\

      Ta & 10  & 21 & 77 & 0.17& 4.4 & 99 &0.96& Refs. \onlinecite{propertiesTa, Tafermivel} \\
    \hline
    \hline
    \end{tabularx}
    }
    \caption{Simulated energy gap ($\Delta$) and kinetic inductance of a 100 nm Nb CPW (central conductor: 10 $\mu m$ wide), with $T_C= 9$ K, $\xi= 40$ nm, $\rho= 7 \mu\Omega\cdot$cm \cite{PropertiesNb}, capped with the metals in Ref. \onlinecite{encapsulation} at $T=0.1 T_{C}$ for Au and at $T=0.1 T_{CS'}$ for the other materials. The values of $\gamma$ and $T_C$ were found or calculated using the data in the cited literature, choosing one of the range of values they could assume. For the case of kinetic inductance, we report the ratio between the capped ($L_{k.cap}$) and uncapped ($L_{k.0}$) CPW, illustrating how changes in resistivity and/or energy gap affect the variation in kinetic inductance.}
    \label{tab:prox}
\end{table*}

\begin{figure}[t]
\subfloat[]
{\includegraphics[width=1.6in]{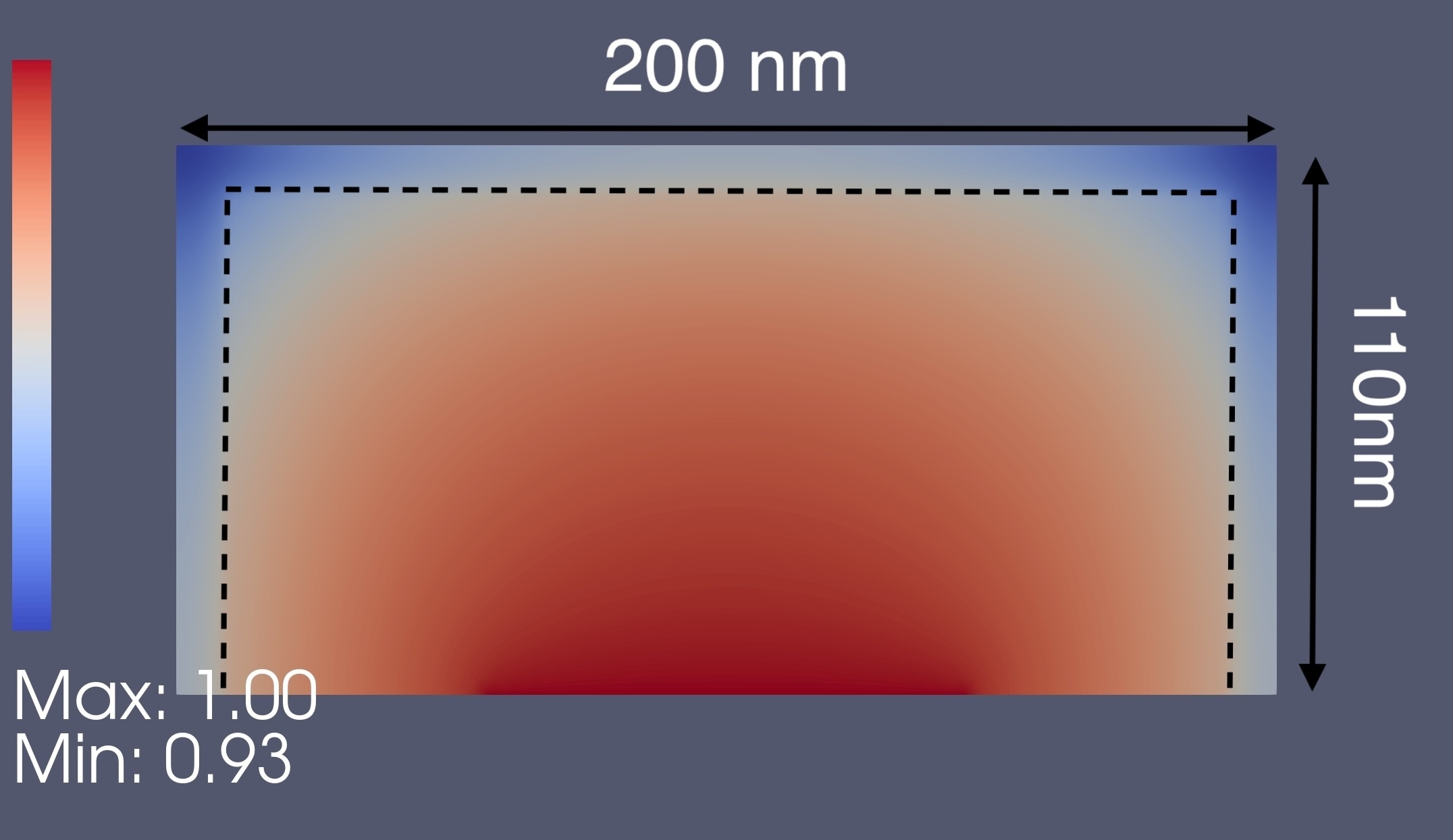}} \quad 
\subfloat[]
{\includegraphics[width=1.6in]{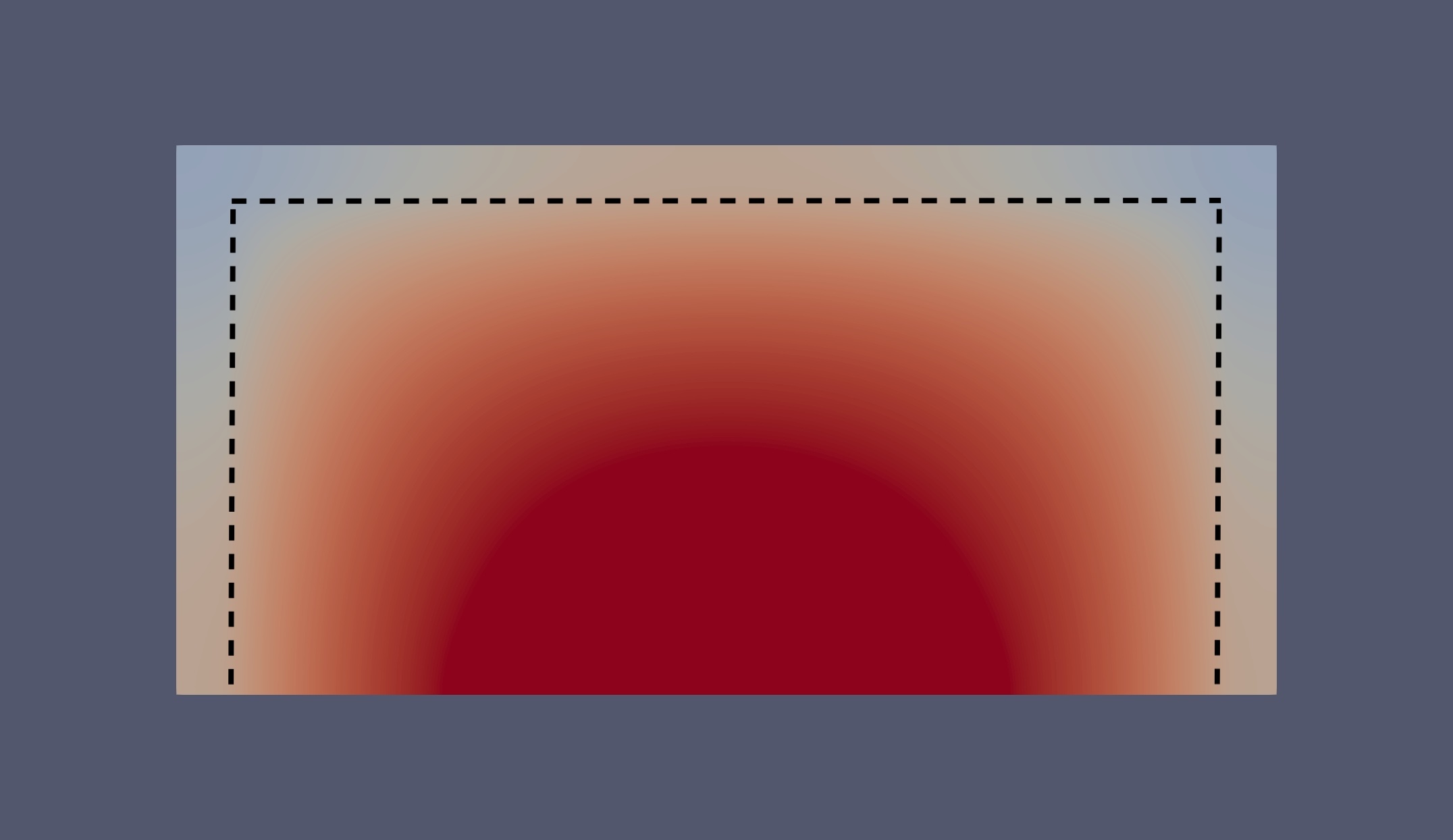}} \\ 
\subfloat[]
{\includegraphics[width=1.6in]{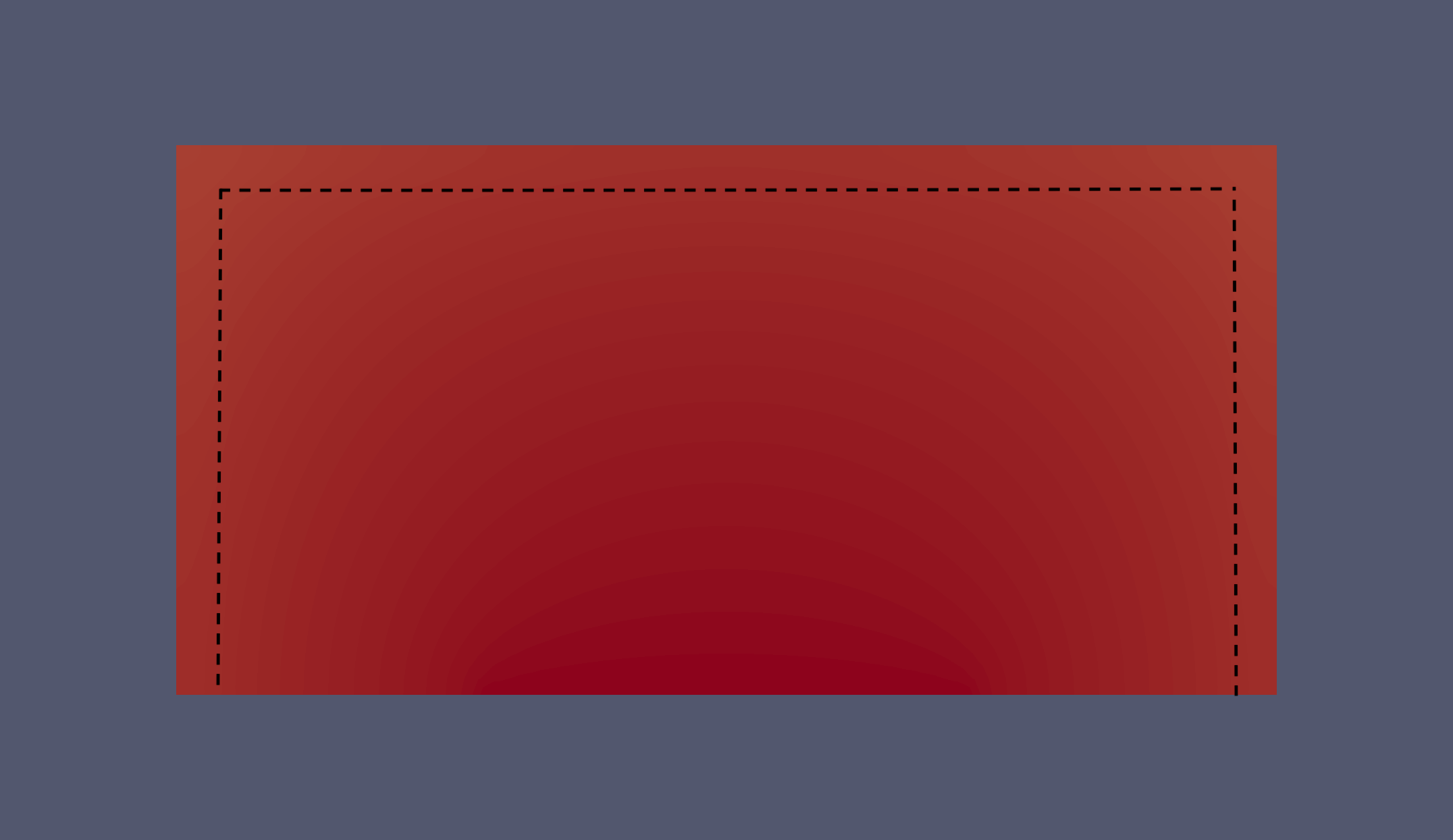}} \quad
\subfloat[]
{\includegraphics[width=1.6in]{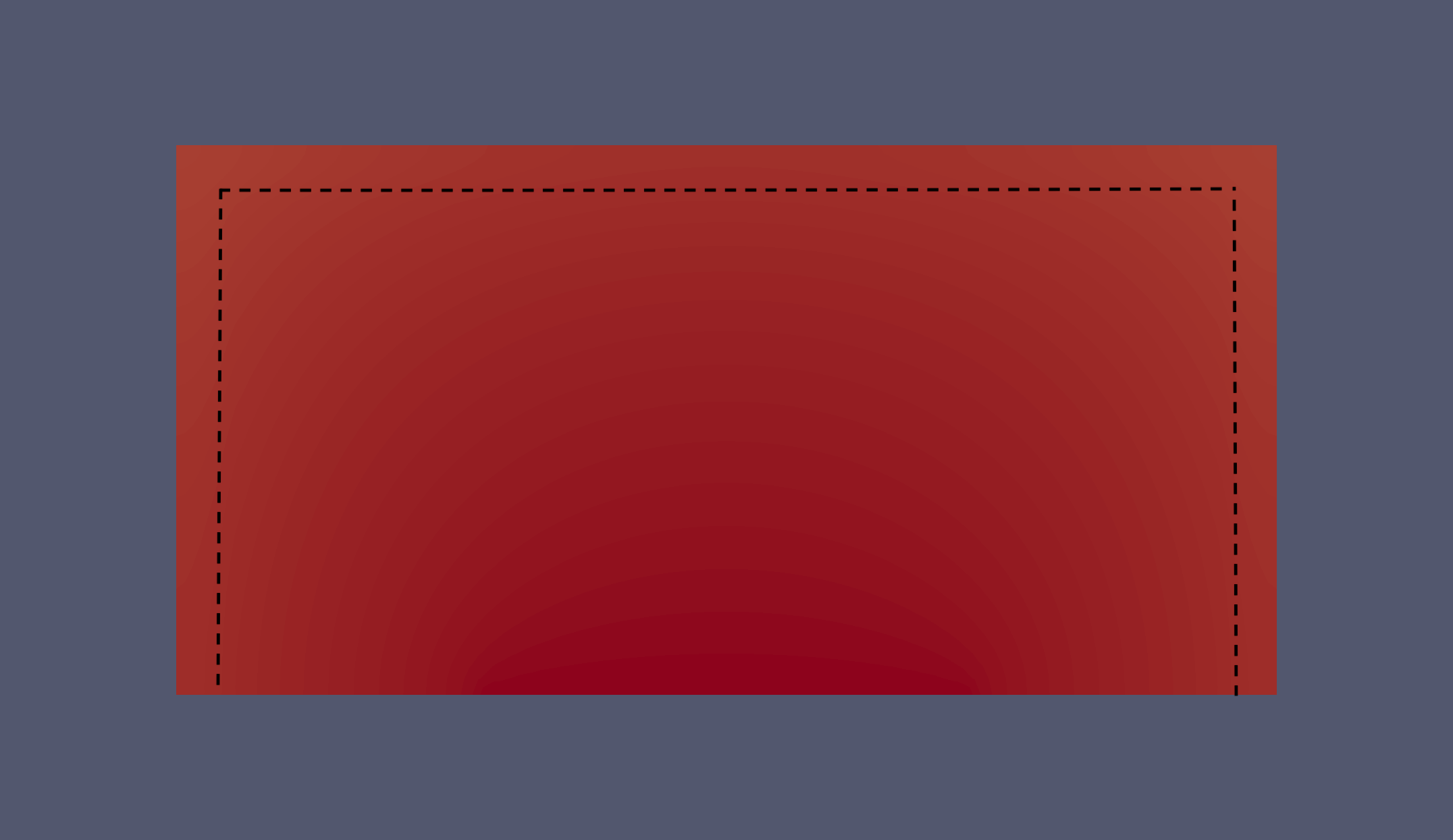}} \quad 
\caption{ Simulated magnitude of the function $\Phi$  in a CPW (Not to scale for better visualization) capped with the materials in Tab. \ref{tab:prox}: a) Au b) Al c) TiN d) Ta. Again, the dotted line was added to the images to indicate the separation between the niobium and encapsulating layer.  Materials with low value of $\gamma$ have low impact on $\Phi$ as expected.}
\label{fig:restab}
\end{figure}

The same equation can be employed in any situation where a superconductor and another metal are in contact. Recently, a method known as encapsulation has been utilized in Ref. \onlinecite{encapsulation} to enhance qubits' lifetime by depositing a normal metal or another superconductor over the superconducting material, in order to mitigate the detrimental effects of surface oxides. However, due to the proximity effect, the energy gap ($\Delta$) decreases in the stronger superconductor and increases in the weaker one causing quasiparticles to start forming at lower drive frequencies. Since it is necessary to satisfy the condition $\hbar f \ll \Delta$ throughout the device \cite{tinkham2004introduction}, they are more susceptible to environmental noise in a cryostat \cite{noisecryo}. This makes it crucial to carefully choose the encapsulating layer for superconducting devices, taking into account the material's thickness and resistivity, as described by the relation $\gamma = \frac{\rho_S\xi_S}{\rho_N\xi_N}$.  In Fig. \ref{fig:res}, we analyze the proximity effect in a superconducting CPW capped with a normal metal, observing that an increase in $\gamma$ leads to a decrease in the superconducting gap, consistent with expectations. In Tab. \ref{tab:prox} and Fig. \ref{fig:restab}, we present calculations of the reduction in $\Delta$ for the materials listed in the referenced paper. This reduction in $\Delta$ leads to an increase in the density of quasiparticles, which in turn reduces the internal quality factor of the resonator and, for qubits, shortens the relaxation time \cite{Decoqubit}. Therefore, to minimize quasiparticle formation and associated losses, it is advisable to select materials with a high critical temperature ($T_C$) and/or values of resistivity and coherence length such that $\gamma$<1, such as tantalum and TiN. Furthermore, the energy gap was also used to calculate the kinetic inductance of the film integrating in parallel the following equation\cite{Annunziata,Kineticmodel}:
\begin{equation}
L_{k}\propto \rho \frac{1}{\Delta} \frac{1}{\tanh{(\Delta /2k_BT)}}.
\end{equation}
The ratio of kinetic inductance with and without the capping layer is reported in Table \ref{tab:prox}. The materials that exhibit the most significant changes in kinetic inductance are those showing the largest variations in their energy gap, as well as aluminium due to its low resistivity, as expected. Our findings indicate that even when the variation in the energy gap is not substantial for the listed materials, resistivity still plays a major role in determining the total film's kinetic inductance, even for small capping-layer thicknesses.

In conclusion, we have developed a detailed numerical model that addresses the nanobridge Josephson junction, coplanar waveguide (CPW) resonator, and qubit as integrated elements in superconducting quantum circuits. Our model overcomes the limitations of earlier approaches, which often focused on simplified geometries or treated these components in isolation. By incorporating realistic 3D geometries, rounded edges, and material considerations, we provide a simulation framework that closely mirrors the actual devices used in experiments. The current-phase relationships (CPRs) derived from our simulations demonstrate how the critical current and energy gap are influenced by the nanobridge's shape and material composition, providing valuable insights into optimizing qubit designs for improved anharmonicity and coherence times. We validated our model by comparing it with experimental data, showing a strong agreement. The flexibility of our model allows it to be applied not only to standard Josephson junctions but also to more complex scenarios like proximity-effect-driven reductions in energy gaps at material interfaces in superconducting CPWs, which are crucial for qubit performance. Overall, our work provides a robust tool for the design and optimization of superconducting qubit circuits. This advancement paves the way for the development of more reliable and scalable quantum devices. Future work will focus on the experimental validation of the devices presented in these designs.
\section*{Supplementary Material}
This section presents additional simulations exploring different combinations of materials forming the nanobridge junction, along with an example demonstrating the convergence of the simulation as the number of mesh elements increases in the 3D model.
\section*{ACKNOWLEDGEMENTS}
This work was supported by the UK Engineering and Physical Sciences Research Council (EPSRC) grant EPIT025746/1 for the University of Glasgow. The authors acknowledge V. Georgiev, L. Kaczmarczyk and A. Shvarts for the fruitful discussions.
\section*{AUTHOR DECLARATIONS}
\subsection*{ CONFLICT OF INTEREST}
The authors have no conflicts of interest to disclose.
\subsection*{ AUTHOR CONTRIBUTIONS}
\textbf{Giuseppe Colletta}: Conceptualization (equal), Software (lead), Methodology (lead), Writing/Original Draft Preparation (lead), Writing – review and editing (equal). \textbf{Susan Johny}:  Conceptualization (equal), Writing – review and editing (equal). \textbf{Jonathan A. Collins}: Conceptualization (equal), Writing – review and editing (equal). \textbf{Alessandro Casaburi}: Conceptualization (equal), Writing – review and editing (equal), Funding Acquisition (equal). \textbf{Martin Weides}: Conceptualization (equal), Writing – review and editing (equal), Funding Acquisition (equal). 
\section*{DATA AVAILABILITY}
The data that support the findings of this study are available from the corresponding author upon reasonable request.
\section*{References}
\bibliography{aipsamp}
\clearpage

\section{Supplementary Material}
\renewcommand{\theequation}{S.\arabic{equation}}
\renewcommand{\thefigure}{S.\arabic{figure}}
\setcounter{equation}{0}
\setcounter{figure}{0}
\subsection{Density of states DOS}
We can also calculate the density of states (DOS) from Eq.(\ref{eq:Usadel}) by an analytical continuation of $\omega_n \rightarrow iE$ and using the value $\Delta(\vec r, T)$ found by solving the Usadel equations before the continuation. Following Ref. \cite{Ragazzidikupr_2021}, we also employed the theta-parametrization of Eq.(\ref{eq:Usadel}) which brings us to this new set of equation:
\begin{equation*}
\label{eq:Dos}
\xi_I^2 \nabla \cdot \bigg (\sin^2 \theta_I \nabla \chi_I \bigg )= \frac{i}{2} \sin \theta_I (\Delta_I e^{-i\chi_I} - \Delta_I^* e^{i\chi_I}),
\end{equation*}
\begin{multline*}
   -\xi_I^2 \nabla^2 \theta_I + \xi_I^2 \sin \theta_I \cos \theta_I (\nabla \chi_I)^2 = \\
   iE \sin \theta_I+ \frac{1}{2} \cos \theta_I (\Delta_I e^{-i\chi_I} + \Delta_I^* e^{i\chi_I}). 
\end{multline*}
and the boundary conditions in Eq.(\ref{eq:boundaryN}, \ref{eq:boundaryN2},  \ref{eq:boundaryext}) become:
\begin{equation}
\cos \theta_S = \frac{E}{\sqrt{E^2 - \Delta_0^2}},
\end{equation}
\begin{equation*}
\chi_S = \pm \frac{\delta}{2}.
\end{equation*}
\begin{equation*}
\gamma_B \xi_{S'/N} \sin^2 \theta_{S'/N} \nabla \chi_{S'/N} \cdot \vec{n} = \sin \theta_S \sin \theta_{S'/N} \sin(\chi_S - \chi_{S'/N}),
\end{equation*}
\begin{multline*}
\gamma_B \xi_{S'/N}\nabla \theta_{S'/N} \cdot \vec{n} = \\
\sin \theta_S \cos \theta_{S'/N} \cos(\chi_S - \chi_{S'/N}) - \cos \theta_S \sin \theta_{S'/N},
\end{multline*}
\begin{equation*}
\gamma_B \xi_S \sin^2 \theta_S \nabla \chi_S \cdot \vec{n} = \gamma \sin \theta_{S'/N} \sin \theta_S \sin(\chi_{S'/N} - \chi_S),
\end{equation*}
\begin{multline*}
\label{eq:boundaryDos}
\gamma_B \xi_S \nabla \theta_S \cdot \vec{n} =\\ 
\gamma \sin \theta_{S'/N} \cos \theta_S \cos(\chi_{S'/N} - \chi_S) - \gamma \cos \theta_{S'/N} \sin \theta_S.
\end{multline*}
\subsection{SNS junction}
\begin{figure} [!ht]
\centering
\sidesubfloat[]
{\includegraphics[width=2.75 in]{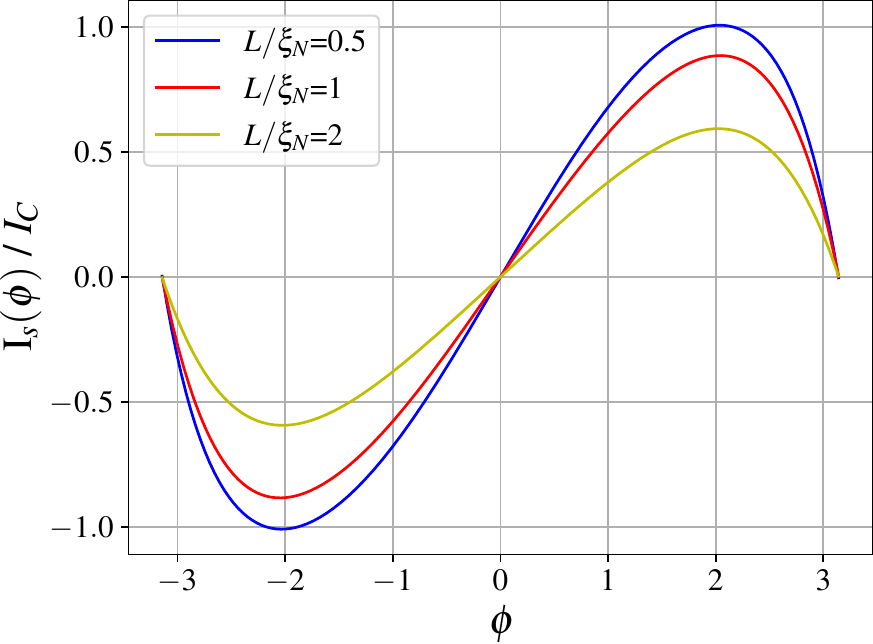}}\quad
\sidesubfloat[]
{\includegraphics[width=3 in]{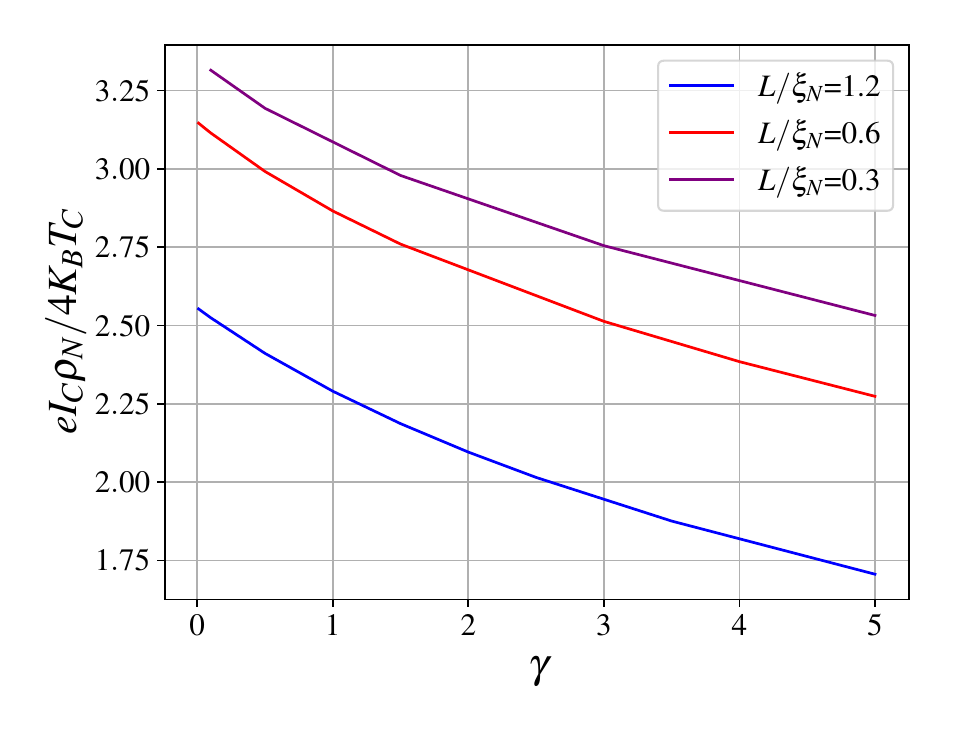}}\\
\caption{(a) Simulated CPR of the VTB nanobridge with rounded cross-section in Fig. \ref{fig:designnano1}, for different values of $\xi_N$, other parameters used were $T/T_C=0.1$, $\xi_S=200$ nm, $L= 60$ nm, $L'=2L$, $\gamma_b=0.01$, $\gamma = 0.01$. (b) Dependence of $I_C$ with $\gamma$ for different values of $\xi_N$, other parameters used were $T/T_C=0.1$, $\xi_S=200$ nm, $L= 60$ nm, $\gamma_b=0.01$}
\label{fig:simulationCPRSNS1}
\end{figure}

Let's consider the case in which the material between the two superconducting banks is a normal metal, i.e. $\Delta=0$. Since these types of junction are characterized by a metallic contact between the two materials, we will consider high transparency interfaces, i.e. $\gamma_b = 0.01$, and different values for $L/\xi$ and $\gamma$. Using the VTB SNS Junction design in Fig.\ref{fig:designnano1}, we obtained the current phase relationships (CPR) in Fig. \ref{fig:simulationCPRSNS1} for different value of $\xi_N$ and then extracted the Critical Currents ($I_C$) for different values of $\gamma$

Then we calculated the density of states in the nanobridge for $L/\xi_N=0.6$, to show at which energies quasiparticles will form in the junction.

\begin{figure} [!ht]
\centering
\includegraphics[width=3 in]{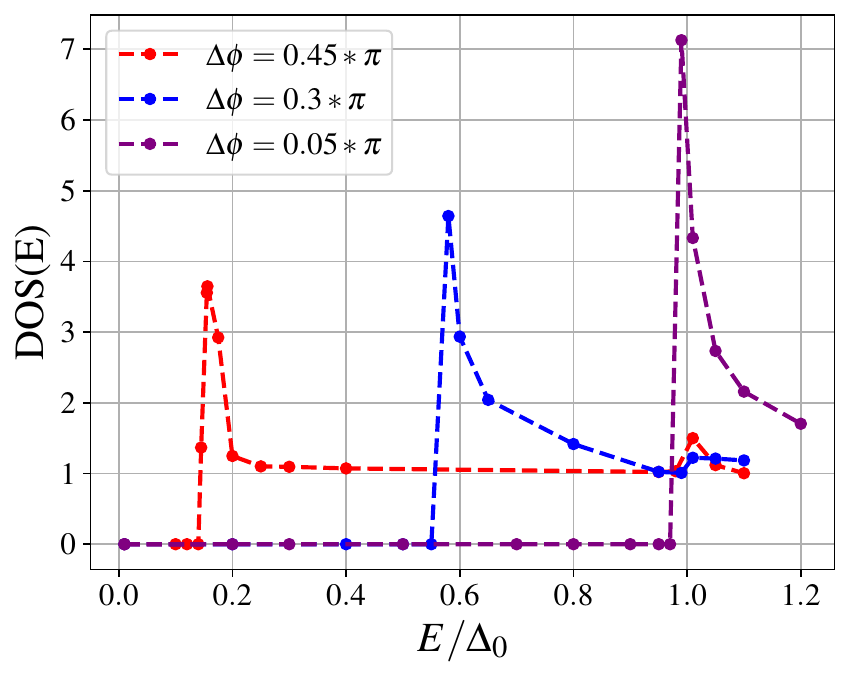}\\
\caption{DOS(E) for a SNS nanobridge junction, calculated in the bridge region for different values of $\Delta \Phi$. Parameters used were $T/T_C=0.1$, $\xi_S=200 nm$, $\xi_N=120 nm$, $W=20$nm, $L= 60$ nm,$L'=2L$, $\gamma_b=0.01$, $\gamma = 0.01$.}
\label{fig:DOSESNS}
\end{figure}

The dependence of the critical current with $\gamma$ and $\xi$ is in line with the original theory and the 2D model in \cite{KUPRIYANOV1981,Ragazzidikupr_2021}. Increasing $\gamma$ or decreasing $\xi_N$ leads to lower values of $I_C$ while not affecting the CPR shape. As we will see in the next chapter this is not the case for SS'S junction where the CPR as a strong dependence on $\xi$.

\subsection{SS'S junction}
\subsubsection{$T_{CS'} > T_{CS}$}
High coherence length materials with high resistivity are needed for superconducting quantum devices since these will correspond to lower critical current and CPR closer to a sinusoidal shape. To obtain junctions with high resistance, one possible solution is to reduce the thickness of the bottom layer to the order of a few tens of nanometers. However, such a reduction adversely affects the quality of the film, leading to negative impacts on critical temperature $T_C$, electronic mean free path ($l$), and, consequently, the material's coherence length. Motivated by these considerations, we investigate a multilayer device, where the lower layer is strategically chosen to possess a higher critical temperature (compared to the top superconductor) and a thickness to obtain the required resistance. Meanwhile, the upper layer is selected to have low $T_C$ and high coherence length. This is because the renormalization of Eq.(\ref{eq:Usadel}) results in an effective coherence length proportional to $\sqrt{\frac{T_{CS'}}{T_{CS}}}$.

This innovative structural configuration not only positively impacts the intrinsic properties of the material but also yields benefits for the surrounding electronics, such as transmission lines. In the nanobridge area, only the lower layer is exposed, while in other regions, the combined superconducting film is preserved. This configuration effectively reduces the risk of producing high kinetic inductance in comparison to scenarios where only the bottom layer is present.

\begin{figure} [!h]
\centering
\includegraphics[width=3 in]{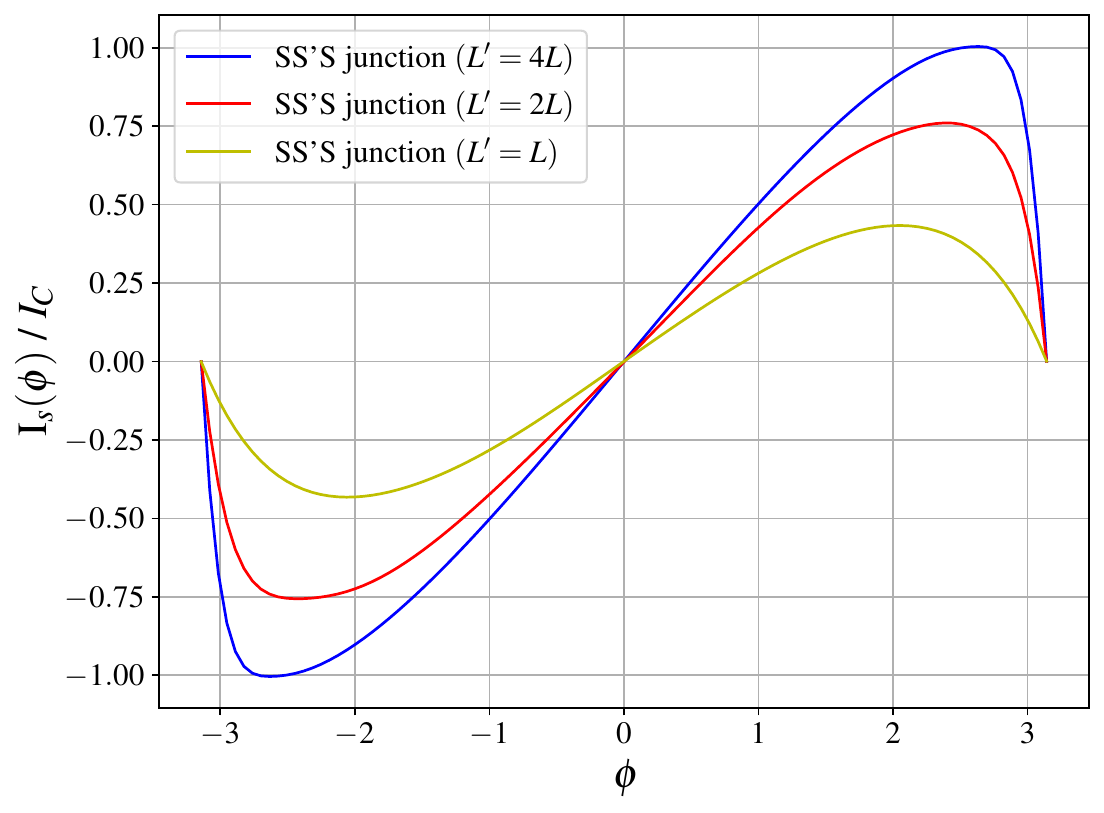}\\
\caption{Simulated CPR of the VTB nanobridge with rounded cross-section in Fig. \ref{fig:designnano1}, other parameters used were $T/T_{CS}=0.1$, $T_{CS}/T_{CS'}=5$, $\xi_S=200 nm$, $\xi_{S'}=30nm$, $W=20 $nm, $L= 60$ nm, $\gamma_b=0.01$, $\gamma = 0.01$. For smaller distances between the banks the CPR turns more sinusoidal.}
\label{fig:simulationCPRA}
\end{figure}

We considered the specific design of VTB SS'S junction with rounded edges in Fig. \ref{fig:designnano1} and conducted simulations for different values of $L/L'$ keeping fixed $\gamma=\gamma_B=0.01$. Other parameters are reported in Fig. \ref{fig:simulationCPRA}.
As we can see in Fig. \ref{fig:simulationCPRA} the CPR gets less and less sinusoidal the more the banks are placed far away from the constriction, becoming more similar to the planar case. This is expected because the S' film has a shorter coherence length than this distance, so it recovers its original properties after a few couples of coherence lengths.

\begin{figure} [!h]
\centering
\includegraphics[width=3 in]{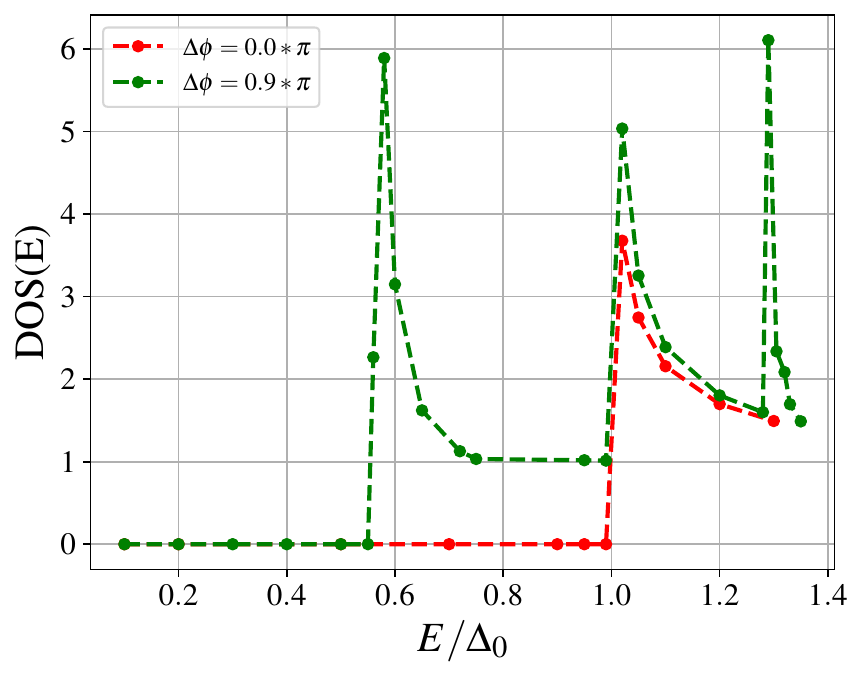}
\caption{Simulated DOS(E) of SS'S junction with $L'=2L$, other parameters are: $T/T_{CS}=0.1$,$T_{CS}/T_{CS'}=5$,, $\xi_S=200 nm$, $\xi_{S'}=30nm$, $W=20 $nm, $L= 60$ nm, $\gamma_b=0.01$, $\gamma = 0.01$. The DOS(E) have a peak near a singularity where quasiparticles start to form, the interesting fact is that with two superconductors more than 2 peaks appear, this corresponds to the formation of quasiparticles due to the presence of the superconductor with a higher gap energy in the bottom layer, so the energy required to generate them is going to be greater than the one for the banks.}
\label{fig:simulationDOSE1}
\end{figure}

We have also calculated the density of states for one of the junction, this quantity is useful to determine the formation of quasi-particles in the junction when an excitation of given energy is applied to it. In Fig. \ref{fig:simulationDOSE1} it can be seen that the effect of using two different layers of superconductors adds a peak at higher energies that comes from the bottom layer having a higher $T_C$.

\section{Convergence of Simulations}  
In this section, we demonstrate the convergence of the simulations as the number of elements in the 3D mesh of the nanobridge model increases. The results show that the critical current exhibits exponential convergence as the number of mesh elements increases.

\begin{figure} [!h]  
\centering  
\includegraphics[width=3 in]{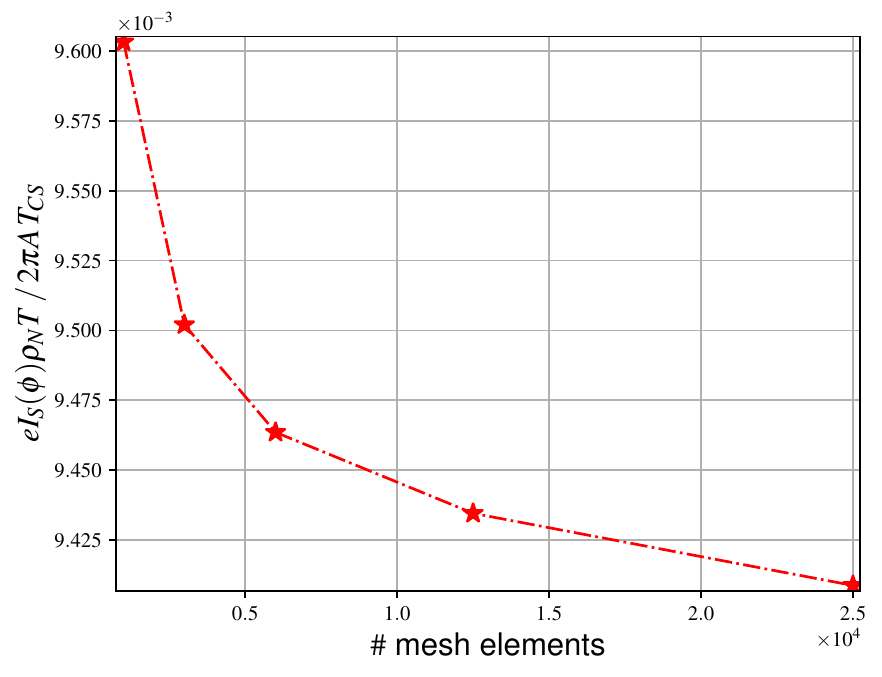}  
\caption{Convergence of the critical current with an increasing number of mesh elements, keeping the number of steps fixed.}  
\label{fig:simulationConv}  
\end{figure}  
\clearpage
\clearpage
\end{document}